\documentclass[baaa]{baaa}

\usepackage[pdftex]{hyperref}
\usepackage{subfigure}
\usepackage{natbib}
\usepackage{helvet,soul}
\usepackage[font=small]{caption}

\usepackage{lipsum} 

\contriblanguage{1}

\contribtype{1}

\thematicarea{7}

\received{\ldots}
\accepted{\ldots}

\title{Connection between feedback processes and the effective yields of EAGLE galaxies}

\titlerunning{Effective yields in EAGLE galaxies}

\author{
M.C. Zerbo\inst{1,2,3}, 
M.E. De Rossi\inst{1,2}, 
M.A. Lara-L\'opez\inst{4,5},
S.A. Cora\inst{3,6} \& 
L.J. Zenocratti\inst{3,6}
}

\authorrunning{Zerbo et al.}

\contact{candelazerbo@gmail.com}

\institute{
Instituto de Astronomía y Física del Espacio, CONICET--UBA, Argentina 
\and
Facultad de Ciencias Exactas y Naturales, UBA, Argentina 
\and
Facultad de Ciencias Astronómicas y Geofísicas, UNLP, Argentina 
\and
Departamento de Física de la Tierra y Astrofísica, Universidad Complutense de Madrid, España
\and
Instituto de Física de Partículas y del Cosmos IPARCOS, Facultad de Ciencias Físicas, Universidad Complutense de Madrid, España
\and
Instituto de Astrofísica de La Plata, CONICET--UNLP, Argentina
}

\resumen{Los mecanismos de {\em feedback} desencadenados por los eventos de supernova (SN, por sus siglas en inglés) y núcleos activos de galaxias (AGN, por sus siglas en inglés) desempeñan un rol central en la regulación de la formación estelar y en el modelado de las propiedades de las galaxias. Sin embargo, cuantificar el impacto y eficiencia de estos procesos continúa siendo un desafío. En este trabajo, utilizamos el conjunto de simulaciones hidrodinámicas cosmológicas {\sc eagle} para examinar diferentes modelos de {\em feedback} de SN y AGN. Nuestro objetivo se centra en estudiar cómo la variación de estos procesos impacta en las propiedades de poblaciones de galaxias simuladas. En este sentido, nos enfocamos en el análisis de los {\em yields} efectivos, evaluando su capacidad de trazar el efecto de los procesos de {\em feedback} en relaciones de escala. Nuestro trabajo contribuye a profundizar la comprensión de la compleja relación que existe entre distintos escenarios de {\em feedback} y la evolución de las galaxias.}

\abstract{The feedback mechanisms triggered by supernova (SN) events and active galactic nuclei (AGN) play a central role on regulating the star formation and shaping galaxy properties. However, quantifying the impact and efficiency of these processes remains a challenge. In this study, we use the {\sc eagle} cosmological hydrodynamics simulations to examine different models of SN and AGN feedback. Our goal is to investigate how variations in these processes impact the properties of simulated galaxy populations. Specifically, we focus on the analysis of effective yields, evaluating their capability to trace the effects of feedback processes on scaling relations. Our work contributes to a deeper understanding of the complex relationship between different feedback scenarios and the evolution of galaxies.}


\begin{document}

\maketitle

\section{Introduction}

The description of the chemical evolution of galaxies inherits the complexity of baryon physics. Initially, the metal enrichment of a galaxy depends on two main aspects: i) the existence of gas reservoirs in which the conditions of temperature and pressure are such that star formation is likely to occur (i.e. star-forming, SF, gas); and ii) the amount of metals released into the interstellar medium (ISM) by a population of stars during their evolution (which is quantified by the nucleosynthetic yields $y_{\rm Z}$). In addition to this, the release of thermal and kinetic energy via supernova (SN) events or the presence of an active galactic nuclei (AGN) modify the conditions of the ISM and, therefore, affect the formation of future generations of stars.

In this article, we focus our study on SN and AGN feedback processes, whose relative efficiencies and impact remain a matter of debate. These processes act via the injection of energy, causing the conversion of SF gas into non-star forming (NSF) material and/or triggering the formation of outflows. 

Historically, observational works have attempted to infer the impact of feedback processes by means of the effective yield, $y_{\rm eff}$. This quantity is calculated from the metallicity of the gas, $Z_{\rm gas}$, and gas mass fraction, $\mu$, of a galaxy in the following way:

\begin{equation}
y_{\rm eff} = \frac{Z_{\rm gas}}{\ln(1/\mu)},
\label{eq:eff_yields}
\end{equation}
where $\mu = M_{\rm gas} / (M_{\rm gas} + M_\star)$, with $M_\star$ and  $M_{\rm gas}$, the stellar and gas masses, respectively. Due to its definition (see \citealt{Dalcanton2007}, for more details), a galaxy that evolves without exchanging mass with its surroundings (referred to as a ``closed-box'') will present a constant value of $y_{\rm eff}$ equal to $y_{\rm Z}$. In contrast, the presence of metal-poor inflows or the formation of metal-enriched outflows will cause the decrease of $y_{\rm eff}$ below $y_{\rm Z}$. 

In this context, \citet[][]{Tremonti2004} reported lower values of $y_{\rm eff}$ in low-mass disk galaxies, which were attributed to the presence of efficient stellar winds. In a more recent study, \citet[][]{Lara-Lopez2019} analysed the relation between $y_{\rm eff}$ and the baryonic mass of galaxies, revealing an anti-correlation above $\sim 10^{10}~\mathrm{M}_\odot$. The authors also compared their results with {\sc eagle} simulations and concluded that AGN feedback is the most likely phenomenon capable of quenching star-formation in massive galaxies, causing the decrease in $y_{\rm eff}$.

In this work, we use {\sc eagle} suite of cosmological hydrodynamical simulations to explore the connection between $y_{\rm eff}$ and feedback processes. Firstly, we examine the properties involved in the definition of $y_{\rm eff}$ and its variations when distinct models of SN and AGN feedback efficiencies are implemented. In particular, we assess the parameter space determined by $M_\star - {\rm O/H} - M_{\rm gas}$, using O/H as a proxy for $Z_{\rm gas}$.
Secondly, we perform an analysis of the physical processes that could drive the trends reported, focusing on the evolution of supermassive black holes (BH) and merger events. Previous results about this project can be found in \citet{Zerbo2022} and  \citet{Zerbo2023}. For a more detailed and extended analysis, the reader is referred to \citet{Zerbo2024}.

\section{The EAGLE simulations}
The {\sc eagle} suite (\citealt{Schaye2015, Crain2015, TheEAGLEteam2017}) is a set of cosmological hydrodynamical simulations that consistently solve the interconnected evolution of dark matter and baryons (i.e. stars, gas and BH). Run in cosmologically representative cubic volumes, all simulations assume a flat $\Lambda$CDM cosmology, with parameters taken from \cite{Planck2015}. Physical processes that can not be directly resolved by equations, are incorporated via subgrid models (i.e. radiative cooling, star formation, SN and AGN feedback, among others).

This set of simulations include ``reference'' models for which free-parameters were selected to reproduce certain scaling relations observed in the Local Universe. Plus, {\sc eagle} provides a collection of runs with variations in one subgrid model parameter. In this work, we assess various SN feedback efficiencies through the comparison of three simulations that include: an attenuated and an enhanced value of efficiency (``WeakFB'' and ``StrongFB'', respectively), and a reference model (``RefL25'') with the same corresponding configuration (i.e. same box length, L, and number of particles per species). In the same way, we analysed AGN feedback effects through the comparison of a model with no AGN physics incorporated (``NoAGN''), a model with a reference value (``RefL50'') and a model where the release of energy is boosted (``AGNdT9''). To obtain further information on these simulations, we recommend consulting \citet{Crain2015}.

\begin{figure}[!t]
\centering
\includegraphics[width=\columnwidth]{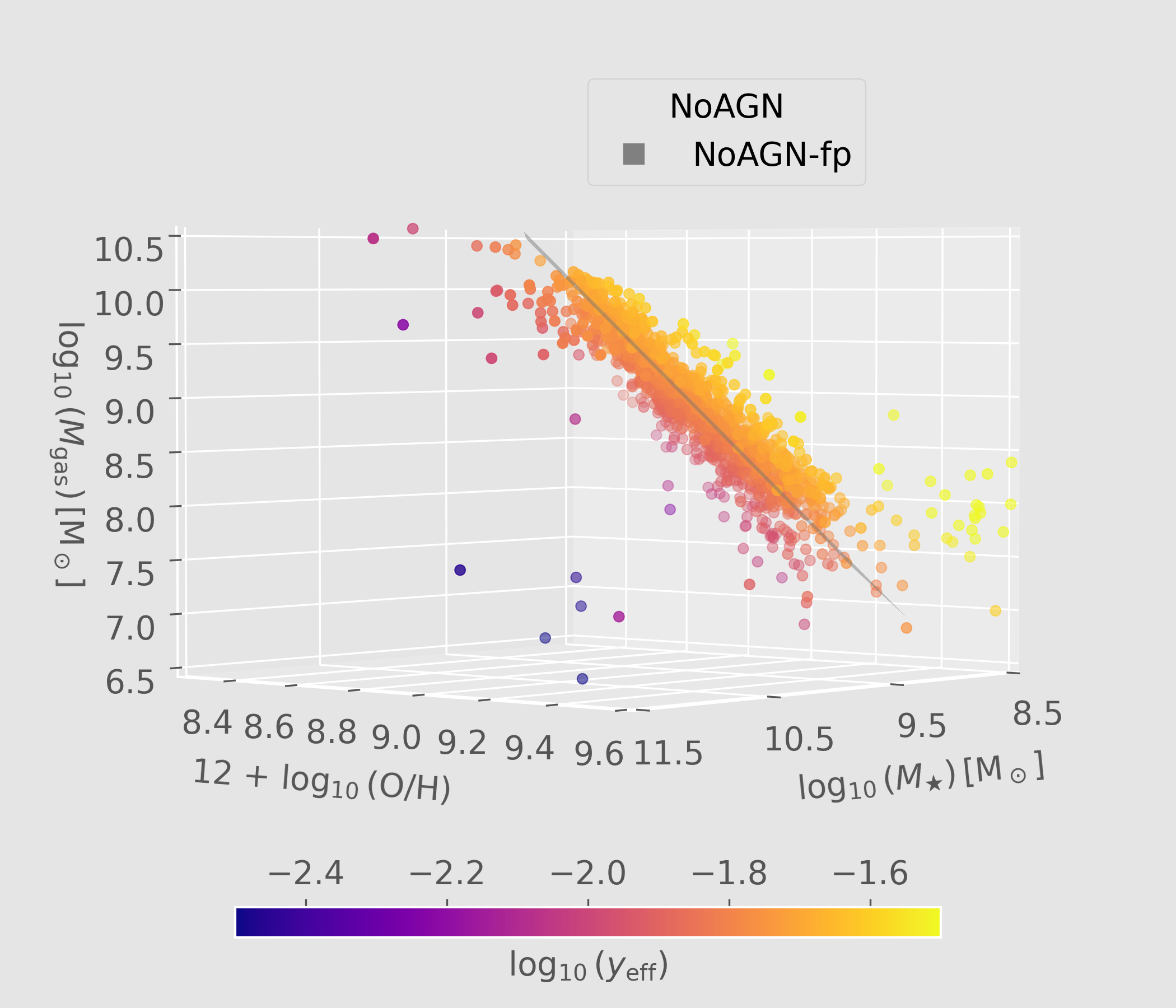}
\includegraphics[width=\columnwidth]{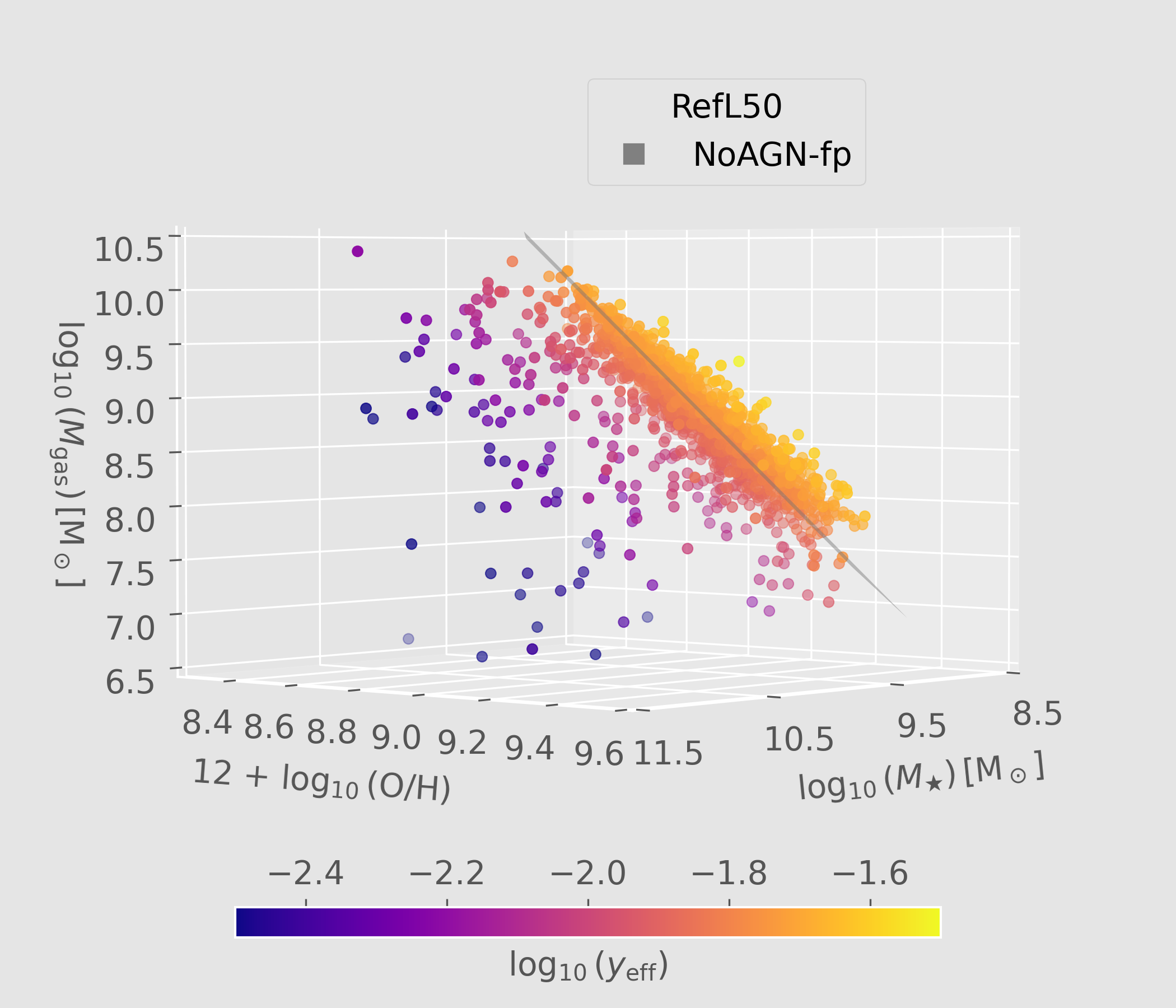}
\caption{$M_\star$ - O/H - $M_{\rm gas}$ relation for ``NoAGN'' (\emph{upper panel}) and `RefL50' (\emph{lower panel}) simulations. Data points are coloured according to $y_{\rm eff}$. The grey line corresponds to the multi-dimensional fit of the ``NoAGN'' data (see Eq.~\ref{eq:plane}).}
\label{Fig:3D_relations}
\end{figure}

\begin{figure*}[!t]
\centering
\includegraphics[width=2\columnwidth]{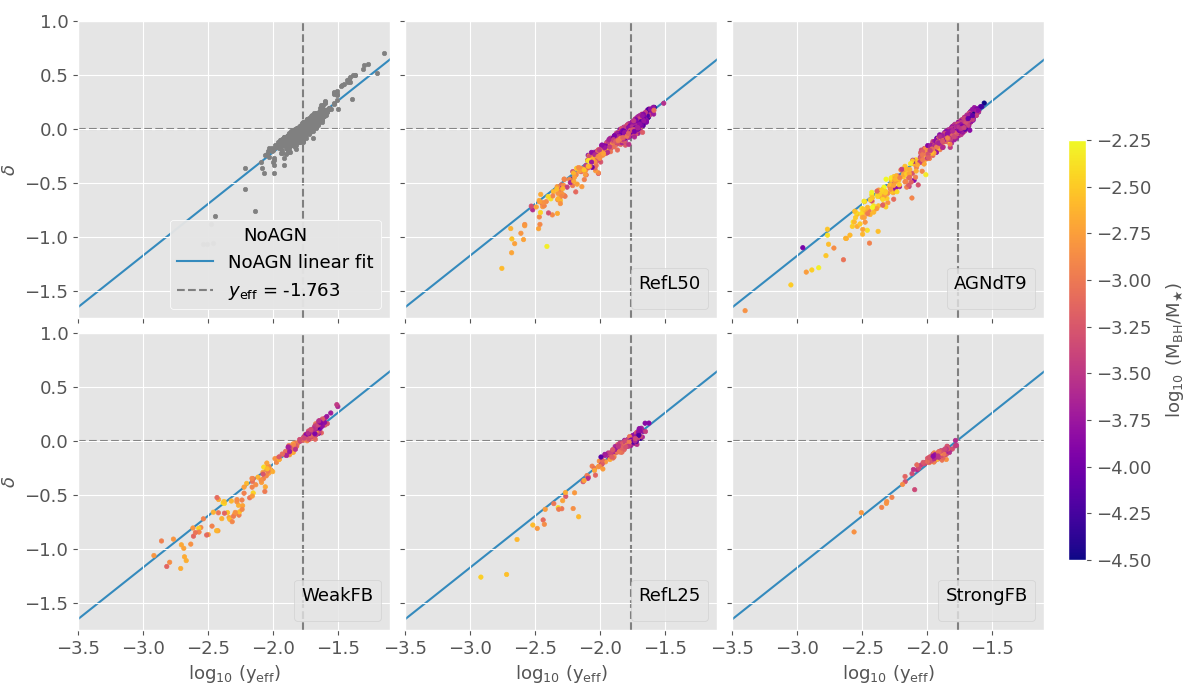}
\caption{Residuals, $\delta$, vs. effective yields, $y_{\rm eff}$, relation for different models. {\em Upper panels}: compare different AGN feedback efficiencies. {\em Lower panels:} assess distinct SN feedback efficiencies. In both cases, the scatter plot is coloured according to the dominance of BH with respect to stellar mass, $M_{\rm BH}/M_\star$. The blue line corresponds to the linear fit performed on the ``NoAGN'' simulation. The dashed vertical line indicates the mean value of $y_{\rm eff}$ in the ``NoAGN'' model. We notice that $M_{\rm BH}$ is null for ``NoAGN'' data, hence galaxies are coloured grey.}
\label{Fig:residue_yeff}
\end{figure*}

\section{Data set and simulated features}
The galaxy features analysed in this work are extracted from the {\sc eagle} database (\citealt{McAlpine2016}). Given that observational studies typically derive measurements from SF regions, we calculate the properties of the gas (e.g. O/H, $M_{\rm gas}$) from the SF component. To prevent numerical artefacts, we limit our sample to galaxies with stellar masses greater than $10^9$~M$_\odot$. While the general trends reported in this work are predominantly traced by central galaxies, this mass threshold also includes a small number of satellites, whose behaviour only contributes to increase the scatter.

In the second part of this work, an statistical evolutionary analysis is carried out by reconstructing the merger trees of galaxies. Therefore, we trace the evolution with time of BH by identifying the main progenitor in each output of the simulation. This particular progenitor satisfies the criteria of being the one with the largest mass considering all earlier simulation outputs. In order to gain more information about merger events, we also identify secondary progenitors, for which the aforementioned condition is not fulfilled. For a detailed description of {\sc eagle} merger trees, the reader is referred to \cite{Qu2017}.

\section{Results}
\subsection{Feedback hints in the $M_\star - {\rm O/H} - M_{\rm gas}$ parameter space}
\label{Sec:3D_relation}
The first part of this work is dedicated to the analysis of the 3-dimensional scale relation that arises in the parameter space defined by $M_\star - {\rm O/H} - M_{\rm gas}$. As reported in \citet{Zerbo2024}, {\sc eagle} galaxies in the ``NoAGN'' model can be well described by a plane, denoted ``NoAGN-fp'' (see Fig.~\ref{Fig:3D_relations}, top panel). This relation follows the expression:

\begin{equation}
    \begin{aligned}
        \log_{10}{\left( \frac{M_\star}{\rm M_\odot} \right)} = \, &C_1 \log_{10}\left( \frac{M_{\rm gas}} {\rm M_\odot} \right) + \\ &C_2 \left[12 + \log_{10} ({\rm O/H})\right]  + C_3, \\
    \label{eq:plane} 
    \end{aligned}
\end{equation}
with $C_1 = 0.844 \, (\pm 0.013)$  , $C_2 = 2.49 \, (\pm 0.05)$ and $C_3 = -20.5 \,(\pm 0.5)$. Taking into account that the axis variables selected are included in the definition of $y_{\rm eff}$ (see eq.~\ref{eq:eff_yields}), when galaxies are colour-coded according to their value of $y_{\rm eff}$, a gradient is formed. In Fig.~\ref{Fig:3D_relations}, top panel, we observe that the ``NoAGN'' model produces a population of galaxies with an homogeneous distribution of values of $y_{\rm eff}$, which exhibit little dispersion. On the other hand, when we consider the more realistic modelling implemented in ``RefL50'' (Fig.~\ref{Fig:3D_relations}, bottom panel), the scatter increases towards lower values of $y_{\rm eff}$ and a fraction of galaxies depart under\footnote{We consider that galaxies are placed under
(over) the plane when, at a certain $M_\star$ and O/H, $M_{\rm gas}$ is lower (higher) than the value given by Eq.~\ref{eq:plane}.} the ``NoAGN-fp''.

To better quantify these deviations, \citet{Zerbo2024} define the residue of a galaxy, $\delta$, as the minimum deviation from the ``NoAGN-fp''. By plotting $\delta$ against $y_{\rm eff}$ (Fig.~\ref{Fig:residue_yeff}), a tight correlation appears in all simulations selected to test SN and AGN feedback efficiency. Furthermore, performing a linear fit to the ``NoAGN'' model provides a relation between $y_{\rm eff}$ and $\delta$ that describes galaxies in all simulations. 

In the top panels of Fig.~\ref{Fig:residue_yeff}, we observe that galaxies evolving with no AGN feedback reach $z=0$ with $y_{\rm eff}$ close to the mean value of $-1.763$. As AGN feedback efficiency increases, a larger number of galaxies present dominant BH and deviate towards lower negative values of residuals (i.e. lower $y_{\rm eff}$). 

In the bottom panels of Fig.~\ref{Fig:residue_yeff}, two effects are observed: a) as SN efficiency increases, the population of galaxies does not reach $y_{\rm eff}$ as high as in other models. This is attributed to the strong effect of winds which create efficient outflows that prevent star-formation. Consequently, the chemical enrichment of galaxies is decelerated, leading to a decrease in the maximum achievable value of $y_{\rm eff}$; b) as SN efficiency decreases, galaxies are capable of developing more dominant BH at all stellar masses. In this scenario, outflows are less likely to form, and more material becomes available for the BH to accrete. With the presence of galaxies with dominant BH, the scatter below the plane increases.

As an overall effect, galaxies with small $M_{\rm BH}$ with respect to their $M_\star$ ($\log_{10}(M_{\rm BH}/M_\star)<-3.5$) remain close to the ``NoAGN-fp'' exhibiting high values of $y_{\rm eff}$. On the other hand, galaxies with dominant BH ($\log_{10}(M_{\rm BH}/M_\star)>-3.5$) depart under the plane, reaching the lowest values of $y_{\rm eff}$.

\subsection{Statistical analysis on evolutionary histories}
\begin{figure}[!t]
\centering
\includegraphics[width=\columnwidth]{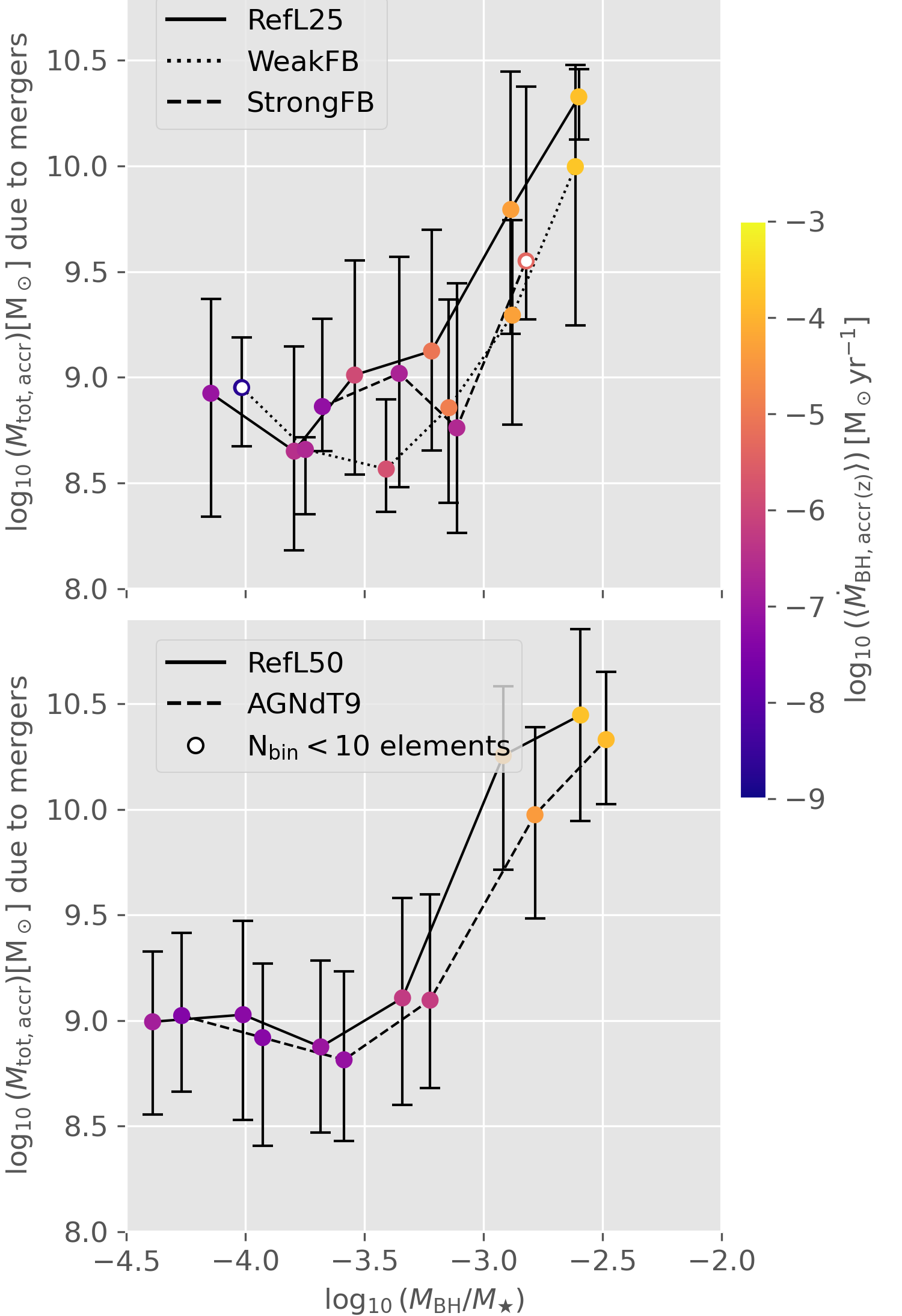}
\caption{{\em Upper panel}: Median relation between the total accreted mass through merger events, $M_{\rm tot, accr}$, and the dominance of BH, $M_{\rm BH}/M_\star$, coloured by the historical accretion rate of BH, $\langle \Dot{M}_{\rm BH}(z) \rangle$, for different SN feedback models. Bins that contain less that 10 galaxies are represented by white circles. {\em Lower panel}: Same as upper panel for different AGN feedback models. No data is displayed for the ``NoAGN'' simulation, since the quantities related to BH are null.}
\label{Fig:evol_histories}
\end{figure}

To obtain a more comprehensive insight regarding the processes that drive the trends observed in Sect.~\ref{Sec:3D_relation} at $z=0$, we study the temporal evolution of BH and the characteristics of merger events. For this analysis, we incorporate two more variables: the historical accretion rate of BH, $\langle \Dot{M}_{\rm BH}(z) \rangle$, which is the median value of the BH accretion rate along the main branch; and the total mass (SF gas + NSF gas + stars) accreted through merger events (i.e. the total mass provided by secondary progenitors), $M_{\rm tot, accr}$.

Fig.~\ref{Fig:evol_histories} shows $M_{\rm tot, accr}$ vs. $M_{\rm BH}/M_\star$ colour-coded by $\langle \Dot{M}_{\rm BH}(z) \rangle$. We notice that galaxies with dominant BH correlate with both larger amounts of total mass accreted via merger events and higher historical rates of BH accretion. These results suggest that merger events could favour the growth of BH. Specifically, disk instabilities can be triggered, causing the migration of material to the inner parts of the galaxy, which becomes more likely to be accreted by BH. As larger amounts of mass are incorporated through mergers, more material is expected to suffer this effect. These trends are visible in all simulations except for ``NoAGN'', for which the quantities related to BH physics are null.
\section{Conclusions}
All things considered, the distribution of values of $y_{\rm eff}$ for a population of galaxies presents distinct features depending on the values of SN and AGN feedback efficiencies. Our findings are in agreement with previous works, which link low values of $y_{\rm eff}$ to galaxies strongly affected by feedback processes. By performing a deeper analysis of simulated galaxies, our results suggest that the trends observed at $z=0$ are a consequence of the accumulated effects of feedback events over the mass assembly histories of galaxies and the evolution of BH. Nevertheless, both feedback processes are interconnected, challenging the idea of isolating the dominant feedback process undergone by a galaxy during its evolution. 
\begin{acknowledgement}
MCZ thanks the Asociación Argentina de Astronomía for providing a grant, which partially support this project. 
We acknowledge support from {\it Agencia Nacional de Promoci\'on de la Investigaci\'on, el Desarrollo Tecnol\'ogico y la Innovaci\'on} (Agencia I+D+i, PICT-2021-GRF-TI-00290, Argentina). 
SAC acknowledges funding from CONICET (PIP-2876), Agencia I+D+i (PICT-2018-3743), and the {\it Universidad Nacional de La Plata} (G11-150), Argentina. We acknowledge the Virgo Consortium for making their simulation data available. The {\sc eagle} simulations were performed using the DiRAC-2 facility at Durham, managed by the ICC, and the PRACE facility Curie based in France at TGCC, CEA, Bruyères-le-Châtel. This work used the DiRAC@Durham facility managed by the Institute for Computational Cosmology on behalf of the STFC DiRAC HPC Facility (www.dirac.ac.uk).
\end{acknowledgement}


\bibliographystyle{899}
\small
\bibliography{899}

\end{document}